\documentclass[a4paper,UKenglish,cleveref]{lipics-v2021}
\usepackage[T1]{fontenc}
\usepackage{pgfgantt}
\usepackage{multicol}
\usepackage{amsmath}  
\usepackage{amssymb}  
\usepackage{graphicx} 
\usepackage{nicefrac}
\usepackage{xspace}
\usepackage{standalone}
\usepackage{tikz}
\usetikzlibrary{patterns}
\usepackage[normalem]{ulem}
\usepackage{todonotes}
\usepackage{cleveref}
\newtheorem{property}[theorem]{Property}
\crefname{property}{Property}{Properties}
\Crefname{property}{Property}{Properties}

\newcommand{\opt}{\textsc{opt}\xspace}
\newcommand{\ceone}{I}

\newcommand{\opti}{\textsc{opt}(\ceone)\xspace}

\newcommand{\lw}{\textsc{lw}\xspace}

\newcommand{\rw}{\textsc{rw}\xspace}
\newcommand{\alg}{\textsc{alg}\xspace}

\newcommand{\ce}{I}

\pgfdeclarelayer{foreground}
\pgfdeclarelayer{background}
\pgfsetlayers{background, main, foreground}
\definecolor{palegray}{gray}{0.8}
\definecolor{paillilac}{RGB}{200,180,255}
\definecolor{lil}{HTML}{AD52A2}
\definecolor{bra}{HTML}{AD8B52}
\definecolor{gru}{HTML}{52AD5D}
\definecolor{bla}{HTML}{365A8C}

\newcommand{\ubratio}{1.273986\xspace}
\newcommand{\ubsimple}{1.274\xspace}
\newcommand{\lbratio}{1.238055\xspace}
\newcommand{\lbsimple}{1.238\xspace}
\newcommand{\generallb}{1.46557\xspace}

\title{Minimizing the Weighted Makespan with Restarts on a Single Machine}
\titlerunning{Minimizing the Weighted Makespan with Restarts}

\author{Aflatoun Amouzandeh}{University of Siegen, Department of Mathematics, Germany}{aflatoun.amouzandeh@uni-siegen.de}{https://orcid.org/0009-0009-6912-3344}{}
\author{Klaus Jansen}{Kiel University, Department of Computer Science, Germany}{kj@informatik.uni-kiel.de}{https://orcid.org/0000-0001-8358-6796}{}
\author{Lis Pirotton}{Kiel University, Department of Computer Science, Germany}{lpi@informatik.uni-kiel.de}{https://orcid.org/0009-0001-4984-3696}{}
\author{Rob {van Stee}}{University of Siegen, Department of Mathematics, Germany}{rob.vanstee@uni-siegen.de}{https://orcid.org/0000-0002-3664-0865}{}
\author{Corinna Wambsganz}{Kiel University, Department of Computer Science, Germany}{cwa@informatik.uni-kiel.de}{https://orcid.org/0009-0002-4820-9017}{}

\authorrunning{A. Amouzandeh, K. Jansen, L. Pirotton, R. van Stee, and C. Wambsganz}
\Copyright{Aflatoun Amouzandeh, Klaus Jansen, Lis Pirotton, Rob van Stee, and Corinna Wambsganz}


\funding{Aflatoun Amouzandeh was financially supported by the Deutsche Forschungsgemeinschaft (DFG, German Research Foundation) - Project Number STE 1727/4-1.
Klaus Jansen, Rob van Stee and Corinna Wambsganz were funded by the Deutsche Forschungsgemeinschaft (DFG, German Research Foundation) -- Project Number 453769249.
Klaus Jansen and Lis Pirotton were funded by the Deutsche Forschungsgemeinschaft (DFG, German Research Foundation) -- Project Number 528381760.}

\relatedversiondetails[cite=AJPSW26]{Preliminary version}
{https://doi.org/10.1007/978-3-032-27732-9_4}

\hideLIPIcs
\nolinenumbers 

\begin{document}
\maketitle

\begin{abstract}
We consider the problem of minimizing the weighted make\-span on a single machine with restarts. 
Restarts are similar to preemptions but weaker: a  job can be interrupted, but then it has to be run again from the start instead of resuming at the point of interruption later. The objective is to minimize the weighted makespan, defined as the maximum weighted completion time of jobs.

We establish a lower bound of \generallb on the competitive ratio achievable by deterministic online algorithms. For the case where all jobs have identical processing times, we design and analyze a deterministic online algorithm that improves the competitive ratio to better than \ubsimple. Finally, we prove a lower bound of \lbsimple for this case. 
\end{abstract}


\section{Introduction}
 \emph{Makespan minimization} is a fundamental and extensively studied problem in scheduling theory, with a significant body of research devoted to it. A sequence of jobs  must be scheduled non-preemptively on $m$ identical machines. Each job $ j $ is characterized by a processing time $ p_j $ and a non-negative arrival time $ r_j $, where $ 1 \leq j \leq n $. In this paper, we assume that jobs arrive over time, i.e., each job becomes available for processing only at its release time. The goal is to minimize the \emph{makespan}, defined as the maximum completion time of any job in the schedule. The problem is obviously NP-hard as it generalizes the makespan problem where all jobs arrive at time 0. Regarding the online setting, the best competitive ratio currently known, for general $m$, is equal to 1.5 and attained by the LPT (Longest Processing Time First) algorithm~\cite{CV97}.

A more general objective is the \emph{weighted makespan}, denoted as $WC_{\max} = \max_j \{ w_j C_j \} $, which measures the maximum \emph{weighted completion time} of all jobs. Each job $ j $ is assigned a positive weight $ w_j $, reflecting its relative importance. The classical makespan minimization problem is a special case in which all weights are equal. This objective function was first considered only fairly recently and has received some attention~\cite{chai2018best,li2015best,FengYuan,sun2024randomized}. 
Even on a single machine, it is nontrivial to optimize the weighted makespan non-preemptively if jobs arrive over time.
In the standard three-field notation introduced by Graham et al.~\cite{graham1979optimization}, this problem is denoted as $1\mid r_j\mid WC_{\max}$.  If all jobs arrive at time 0 or preemptions are allowed, then the algorithm Largest Weight (LW) which always runs the job that has the largest weight is 1-competitive~\cite{chai2018best,li2015best}.

In the \emph{offline} setting, all job parameters (processing times, release times, and weights) are known in advance, allowing the algorithm 
to make better decisions. In contrast, the \emph{online setting} assumes that jobs arrive over time and decisions must be made without knowledge of the future input. Once a job arrives, the online algorithm learns its parameters and must decide whether to process it immediately or delay its execution. The effectiveness of an online algorithm is measured by its \emph{competitive ratio}~\cite{DBLP:books/daglib/0097013,fiat1998online,DBLP:series/txtcs/Komm16}. For a given instance $ I $, the objective values of the online and optimal offline algorithms are denoted as $\alg(I) $ and $\opt(I) $, respectively. The competitive ratio $ \mathcal{R} $ of an online algorithm is defined as:
\[
\mathcal{R} = \sup_I \frac{\alg(I)}{\opt(I)}.
\]
This ensures that, in the worst case, the online algorithm's cost does not exceed $\mathcal{R}$ times the optimal cost. An online algorithm may use as much computation as needed to make decisions, but it lacks knowledge of the input. 

Feng and Yuan~\cite{FengYuan} were the first to consider the weighted makespan in the context of single-machine scheduling. Li~\cite{li2015best} explored the online version of this problem and established a lower bound of 2 on the competitive ratio for deterministic algorithms. Moreover, for the problem on a single machine, he  provided an online algorithm with a competitive ratio of 3.  For the case of identical machines where all jobs have the same processing time, 
Li proposed an online algorithm with the best-possible competitive ratio of $\frac{\sqrt{5}+1}{2}$.

Chai et al.~\cite{chai2018best} further improved the results on a single machine, by developing two deterministic online algorithms with a competitive ratio of 2. Lu et al.~\cite{lu2021single} considered the offline single-machine problem with job rejection but without release dates. They showed that this problem is NP-hard and proposed a fully polynomial-time approximation scheme (FPTAS).  More recently, Sun~\cite{sun2024randomized} proposed a $(2-\frac{1}{m})$-approximation algorithm for the weighted makespan scheduling problem on $m$ identical machines, again in the setting without release dates ($P||WC_{\max}$). The algorithm runs List Scheduling on the jobs after sorting them by decreasing weight. Sun also introduced a randomized efficient polynomial-time approximation scheme (EPTAS) for this problem, as well as a randomized FPTAS for the case when $m$ is fixed. 

One important extension to the classical scheduling framework is the use of \emph{restarts}, which forms a central focus of this paper. In this model, a running job can be interrupted when a more urgent job arrives (e.g., a smaller or heavier job in the case of weighted jobs). However, any work done on the interrupted job is lost, and it must be restarted from the beginning. This model, also known as \emph{preemption with restarts}, differs from the more commonly studied preemption with resume model, where interrupted jobs can be resumed from the point of interruption. The concept of restarts was first introduced by Shmoys et al.~\cite{DBLP:conf/dimacs/ShmoysWW91} in the context of makespan minimization and was later shown to improve the performance of online algorithms for other objectives on a single machine ~\cite{DBLP:conf/esa/AkkerHV00,DBLP:conf/waoa/AmouzandehS24,DBLP:journals/anor/LaalaouiM25,DBLP:journals/jal/SteeP05}, including for equal-length jobs~\cite{DBLP:journals/siamcomp/ChrobakJST07}. However, for the important objective function of total weighted flow time, where the flow time of a job is the time between its arrival and its completion, it is known that restarts do not help~\cite{DBLP:journals/tcs/EpsteinS03}: any online algorithm that can restart jobs has a competitive ratio of $\Omega(n)$. 

Variants of the restart model, including \emph{limited restarts}, which means that  each job may be restarted at most once, and $k$-limited restart where each job can be restarted at most $k$ times and $k$ can be either a positive integer or infinity, have also been studied, particularly in the context of parallel batch machines~\cite{liu2022online,fu2010online,tian2014online}. 

Liang et al.~\cite{liang2024online} recently explored the weighted makespan problem in the restart model for online single-machine scheduling. They proved that when only a single restart is allowed across all jobs, no online algorithm can achieve a competitive ratio lower than 2, matching the lower bound for the problem without restarts~\cite{li2015best}. However,  in the case where all jobs have unit processing times, they presented the best possible online algorithm with a competitive ratio of approximately \generallb. The best known upper bound on the competitive ratio for deterministic online algorithms in the general setting with restarts is $2$ resulting from~\cite{chai2018best}.

In this paper, we consider the problem of online scheduling with restarts for minimizing the weighted makespan $ WC_{\text{max}}$. Specifically, we study the problem $1|r_j, {\text{online}}, {\text{restart}}|WC_{\max}$ and establish a new lower bound of \generallb on the competitive ratio for deterministic online algorithms in the general setting with restarts. Remarkably, this is the same value that Liang et al.~\cite{liang2024online} achieved for jobs of unit size and a single allowed restart. However, the problems are different. 

Our main result is a deterministic online algorithm that achieves a competitive ratio better than \ubsimple for the case where all jobs have identical processing times. 
The algorithm balances a greedy approach, favoring heavier jobs, with the consideration that not too much time should be wasted on jobs that are eventually interrupted anyway. If a job arrives that is sufficiently heavier (by a predetermined factor) than the currently running job, the new job is started instead unless the current job has already been running for sufficiently long. What sufficiently long is depends on the current time: we do not interrupt the running job in the case that the new job, if it is started after the current job completes, still completes within a factor of \ubsimple of its smallest possible completion time (which is achieved if the new job starts immediately when it arrives).

We also prove a lower bound of \lbsimple for this case. 
In determining both lower bounds, we first identified some difficult inputs and then used a computer to optimize the parameters. For the optimized values we then considered in which cases the goal competitive ratio was achieved.
This resulted in a system of equalities that we solved to get exact lower bounds.


\section{Lower bound}\label{sec:generallb}
We now consider the online problem $1|r_j, \text{online} ,\text{restart} | W C_{\max}$. The lower bound presented here shows that the problem maintains its difficulty, even when unlimited restarts are allowed.
\begin{theorem}
Every deterministic online algorithm for
$1|r_j,\text{online},\text{restart}|WC_{\max}$ has competitive ratio
at least $R_1\approx 1.465571$, where $R_1$ is the unique real root of
$x^3-x^2-1=0$. 
\end{theorem}

\begin{proof}
Since $1<R_1<3/2$, we have $0<r:=R_1-1<1/2$ and $w:=R_1^2>1$.
The defining equation for $R_1$ gives
\[
    wr=1,\qquad 1+r=R_1,\qquad 1+w=R_1w.
\]
Fix $0<\varepsilon<1/4$. We give an adaptive instance on which any
deterministic online algorithm $\alg$ has ratio at least
$R_1/(1+2\varepsilon)$. Letting $\varepsilon$ tend to zero proves the theorem.
At time $0$, release job 1 with processing time $p_1=1$ and weight $w_1=1$.
At time $r$, release job 2 with processing time $p_2=\varepsilon$ and
weight $w_2=w$. 

\proofsubparagraph{Case 1: Job 2 has not completed before time $1$.}
No further jobs arrive. Then $\alg(I)\ge w$. An offline schedule runs
job 2 during $[r,r+\varepsilon]$, followed by job 1, so
\[
    \opt(I)\le
    \max\{w(r+\varepsilon),r+\varepsilon+1\}
    =R_1+\varepsilon.
\]
Indeed, $w(r+\varepsilon)=1+w\varepsilon\le R_1+\varepsilon$,
since $(w-1)\varepsilon=(R_1-1)(R_1+1)\varepsilon<R_1-1$.
Consequently,
\[
    \frac{\alg(I)}{\opt(I)}
    \ge\frac{w}{R_1+\varepsilon}
    \ge\frac{R_1}{1+2\varepsilon}.
\]

\proofsubparagraph{Case 2: Job 2 completes before time $1$.}
At time $1$, release job 3 with $p_3=\varepsilon$ and $w_3=w$, and
job 4 with $p_4=w-1$ and $w_4=1$. No further jobs arrive.
An offline schedule runs job 1 during $[0,1]$, followed consecutively
by jobs 2, 3, and 4. Its weighted makespan is $w(1+2\varepsilon)$, hence
$
    \opt(I)\le w(1+2\varepsilon).
$
Write $C_i$ for the completion time of job $i$ in the online schedule.
Since $C_2<1$,
\[
    C_1\ge C_2+1\ge r+\varepsilon+1=R_1+\varepsilon.
\]

If $C_3<C_1$, the final attempt of job 1 also starts at or after job 3
completes. Therefore the successful processing of jobs 1, 3, and 4
all takes place at or after time $1$. Their total processing time is
$1+\varepsilon+(w-1)=w+\varepsilon$, and all their weights are at least $1$.
It follows that
$
    \alg(I)\ge 1+w+\varepsilon\ge R_1w.
$

Otherwise $C_1<C_3$. The final attempt of job 3 starts at or after job 1
completes, so $C_3\ge C_1+\varepsilon\ge R_1+2\varepsilon$ and
$
    \alg(I)\ge wC_3\ge w(R_1+2\varepsilon)\ge R_1w.
$

In either case,
\[
    \frac{\alg(I)}{\opt(I)}
    \ge\frac{R_1w}{w(1+2\varepsilon)}
    =\frac{R_1}{1+2\varepsilon}.
\]
Thus every competitive ratio strictly smaller than $R_1$ is ruled out
by choosing $\varepsilon$ sufficiently small. 
\end{proof}

\section{Unit-sized Jobs}
In this section, we consider the problem of online scheduling with restarts for minimizing the weighted makespan, assuming all jobs have identical processing times. In this case, the problem is equivalent to considering unit-sized jobs, by scaling the release times accordingly. We first present our deterministic online algorithm Relative Weight (\rw) which has a competitive ratio of $R\approx\ubratio$, the  
real root of $6x^5-5x^4-x^3-x^2-x-2=0$.
We also provide a lower bound for this problem.

The idea of our algorithm is as follows. In principle, we prefer to run the heaviest job first. The algorithm which always does this, interrupting other jobs if needed, is called \lw. There are two clear disadvantages to \lw. A job can be interrupted even though
\begin{itemize}
	\item it is nearly finished, or
	\item the new job has only slightly higher weight.
\end{itemize}
We restrict interruptions to avoid both of these bad cases. Regarding the first problem, we consider the \emph{relative completion time} of the incoming job. That is, if the running job $x$ started its current attempt at $t$, an arriving job $y$ at time $u$ interrupts it only if $(t+2)/(u+1)>R$, meaning that if we do not interrupt $x$, then $y$ possibly (depending on the actual optimal schedule) completes more than $R$ times later than in the optimal schedule. The condition is equivalent to $u<(t+2)/R-1$.

\paragraph*{Algorithm RW} Let 
\begin{equation}
	\label{eq:tprime}
	q(t)=\frac{t+2}R - 1.
\end{equation} 
Our algorithm \rw starts a heaviest waiting job whenever the machine becomes free, and does not idle while a job is waiting. If the running job $x$ started its current attempt at $t$, an arriving job $y$ at time $u$ interrupts it precisely when $u<q(t)$ and $w_y>\Delta w_x$. 

We have 
\begin{align*}
	t < \frac{t+2}{R}-1  
	\Leftrightarrow tR -t < 2-R 
	\Leftrightarrow t< \frac{2-R}{R-1}=:T.
\end{align*}
We set $\Delta=\sqrt[4]{(3+3/R)/4}\approx1.076$.

\begin{figure}[h]
    \centering
    \begingroup
    \definecolor{algFigGray}{gray}{0.8}
    \definecolor{algFigLilac}{HTML}{AD52A2}
    \definecolor{algFigBlue}{HTML}{365A8C}

    \def\algFigR{1.274}
    \def\algFigRTwo{0.2}
    \def\algFigRThree{0.8}
    \def\algFigRFour{1.4}
    \pgfmathsetmacro{\algFigQZero}{2/\algFigR-1}
    \pgfmathsetmacro{\algFigCTwo}{\algFigRTwo+1}
    \pgfmathsetmacro{\algFigQTwo}{(\algFigRTwo+2)/\algFigR-1}
    \pgfmathsetmacro{\algFigQFour}{(\algFigRFour+2)/\algFigR-1}
    \pgfmathsetmacro{\algFigJobTwoLabel}{(\algFigQTwo+\algFigCTwo)/2}

    \resizebox{\columnwidth}{!}{%
    \begin{tikzpicture}[
        x=10cm,
        y=1cm,
        every node/.style={font=\large}
    ]
        \fill[algFigGray]
            (0,0.12) rectangle (\algFigQZero,1.12);
        \draw[line width=1pt]
            (0,0.12) rectangle (1,1.12);
        \node[font=\Large] at (0.5,0.62) {Job 1};

        \draw[thick,->] (0,0) -- (1.93,0)
            node[anchor=north] {time};

        \foreach \algFigX/\algFigTickText in {
            0/{t=0},
            \algFigRTwo/{r_2},
            \algFigRThree/{r_3},
            1/{1},
            \algFigRFour/{r_4}
        } {
            \draw (\algFigX,0.035) -- (\algFigX,-0.07)
                node[anchor=north] {$\algFigTickText$};
        }

        \draw (\algFigQZero,0.035) -- (\algFigQZero,-0.43)
            node[anchor=north] {$q(0)=2/R-1\approx
                \pgfmathprintnumber[
                    fixed,precision=3,fixed zerofill
                ]{\algFigQZero}$};

        \begin{scope}[yshift=-3cm]
            \fill[pattern=north east lines,pattern color=algFigLilac]
                (0,0.12) rectangle (\algFigRTwo,1.12);
            \draw[line width=1pt,dash pattern=on 3pt off 2pt]
                (0,0.12) rectangle (\algFigRTwo,1.12);
            \node[font=\Large]
                at ({\algFigRTwo/2},0.62) {Job 1};

            \fill[pattern=north east lines,pattern color=algFigBlue]
                (\algFigRTwo,0.12) rectangle (\algFigQTwo,1.12);
            \draw[line width=1pt]
                (\algFigRTwo,0.12) rectangle (\algFigCTwo,1.12);
            \node[font=\Large]
                at (\algFigJobTwoLabel,0.62) {Job 2};

            \fill[pattern=north east lines,pattern color=algFigLilac]
                (\algFigCTwo,0.12) rectangle (\algFigRFour,1.12);
            \draw[line width=1pt,dash pattern=on 3pt off 2pt]
                (\algFigCTwo,0.12) rectangle (\algFigRFour,1.12);
            \node[font=\Large]
                at ({(\algFigCTwo+\algFigRFour)/2},0.62) {Job 3};

            \fill[algFigGray]
                (\algFigRFour,0.12) rectangle (\algFigQFour,1.12);
            \foreach \algFigY in {0.12,1.12} {
                \draw[line width=1pt]
                    (\algFigRFour,\algFigY) -- (1.8,\algFigY);
                \draw[line width=1pt,dash pattern=on 3pt off 2pt]
                    (1.8,\algFigY) -- (1.9,\algFigY);
            }
            \node[font=\Large] at (1.66,0.62) {Jobs 4,3,1};

            \draw[thick,->] (0,0) -- (1.93,0)
                node[anchor=north] {time};

            \foreach \algFigX/\algFigTickText in {
                0/{0},
                \algFigRTwo/{0.2},
                \algFigRThree/{r_3},
                \algFigRFour/{t=r_4},
                \algFigQFour/{t'}
            } {
                \draw (\algFigX,0.035) -- (\algFigX,-0.07)
                    node[anchor=north] {$\algFigTickText$};
            }

            \draw (\algFigQTwo,0.035) -- (\algFigQTwo,-0.43)
                node[anchor=north] {$2.2/R-1$};
            \draw[dashed]
                (\algFigRTwo,0) -- (\algFigRTwo,1.5);
            \draw[dashed]
                (\algFigRFour,0) -- (\algFigRFour,1.5);
        \end{scope}
    \end{tikzpicture}%
    }
    \endgroup
    \caption{This example shows the behavior of our algorithm.}
    \label{fig:algbehav}
\end{figure}

For any job that starts at some time $t<T$, $q(t)$ is called the \emph{cutoff} of the job. If an interruption occurs, $t$ and $q(t)$ are updated.
In Fig.~\ref{fig:algbehav}, we sketched in an example the behavior of the algorithm for the
following job sequence: 
    $((0,1),(0.2, 1.1), (0.8, 1.6), (1.4, 2.3))$. 
    Here each pair specifies first a release date and then the weight of the job released at this time.
The top figure shows the situation at time $t=0$ and the time until the first cutoff
with values $t$ and $q(t)$. Job $1$ gets interrupted because job $2$ is more than a factor of $\Delta$ heavier. The bottom figure shows the schedule of the algorithm at time $t=r_4$. Job $2$ is not interrupted before the cutoff, so it continues without interruption. At its completion time the algorithm starts the waiting job $3$. Job $3$ gets interrupted before its cutoff and after that the remaining jobs are scheduled in the order $4,3,1$ without further interruptions.

\subsection*{Analysis}
We now prove that \rw is (exactly) $R$-competitive. 
Recall that $R$ is the real root of $6x^5-5x^4-x^3-x^2-x-2=0$.

We call a job \emph{critical} for an input $I$ if its weighted completion time is equal to $\rw(I)$. That is, no other job in the schedule of $\rw$ for input $I$ has a higher weighted completion time. An input $I$ may contain multiple critical jobs.

We assume the existence of a \emph{minimal counterexample} $\ce$. A minimal counterexample has ratio strictly larger than $R$ and consists of the smallest number of jobs. We show by contradiction that such a counterexample does not exist. Without loss of generality, we can assume that no two jobs arrive at the same time, as otherwise, we could postpone the lighter job(s) by an arbitrarily small amount.

Consider an arbitrary critical job in $\ce$ and let $S$ be the maximal set of jobs that run consecutively without interruptions and that ends with this critical job. We scale the input such that this critical job has weight 1. Thus, $\rw(I)$ is simply the completion time of the critical job. 

We now derive a sequence of properties that a counterexample must have, which will eventually lead to the conclusion that no counterexample exists. We begin by showing that the critical job is the lightest job in $S$ (Property \ref{prop:lightest}), and that \opt starts running its first job in $S$ relatively early (Property \ref{prop:early}). \rw starts its first job in $S$ (which we will call job 1, see text below Property \ref{prop:lightest}) when it arrives (Property \ref{prop:1}), and the arrival of job 1 actually caused the interruption of another job $h$ (Property \ref{prop:h}).
We then show that \rw does not complete any job before job 1 (Property \ref{prop:nocomp}), and consider the structure that remains (Theorem \ref{thm:rw}).

\begin{observation}
	\label{obs:simple}
	When \rw starts a job, no heavier job is waiting.
	When a job arrives, \rw does not start a lighter job until it completes.
	When a job is interrupted at time $u$, a heaviest job that arrives at time $u$ is started.
\end{observation}
\begin{proof}
	The first claim is immediate from the algorithm. The second claim is an immediate consequence of the first claim.
	
	Suppose a job starts at time $t$. By the first claim, every waiting job at time $t$ is no heavier than the running job. Before its cutoff, any subsequently arriving job that did not cause an interruption has weight at most \(\Delta\) times its weight. An interruptor at time $u$ exceeds that threshold and is therefore heavier than every waiting job.
	\qedhere 
\end{proof}

\begin{observation}
	\label{obs:running}
	If \rw runs a job $g$ in some interval $[a,b)$, where $g$ started at time $a$, then no job of weight more than $\Delta w_g$ arrives in $(a,\min(b,q(a)))$, and any such job that arrives in $[0,a]$ is completed by time $a$. If a job $p$ is waiting at time $b$ and $w_p>w_g$, then $r_p\ge q(a)$ or $w_g\ge w_p/\Delta$ holds.
\end{observation}
\begin{proof}
	A job of weight more than $\Delta w_g$ would cause an interruption of $g$ if it arrived in $(a,\min(b,q(a)))$. If it arrives earlier, it must also complete earlier by Observation \ref{obs:simple}. Regarding the last claim, job $p$ arrived after time $a$ by Observation \ref{obs:simple} but did not cause an interruption, so we have $r_p\ge q(a)$ or $w_g\ge w_p/\Delta$.\qedhere
\end{proof}

We define $n_0(t)=t$ and $n_m(t)=q(n_{m-1}(t))$ for $m\ge1$. We have $t-T = (t-T)/R^0$ 
and by induction
\[
n_m(t)-T = q(n_{m-1}(t))-q(T) = \frac{n_{m-1}(t)-T}{R} = \frac{t-T}{R^m}
\]
for $m\ge1 $.
Hence 
\begin{equation}
	\label{eq:nmt}
	n_m(t) = T + \frac{t-T}{R^m} = (1-R^{-m})T+R^{-m}t.
\end{equation}
We will also use that $q(t+1)=q(t)+1/R$.

\begin{lemma}
	\label{lem:s1bound}
	Let $0\le a<s_1$. Suppose $(a,s_1]$ is an interval in which \rw is not idle, and suppose \rw starts a job at time $s_1$.
	Suppose the interval contains $\ell$ interruptions and $c$ completions, with $d_i$ interruptions (in total) following the $i$th completion. Then
	\begin{equation}
		\label{eq:events}
		s_1\le n_\ell(a)+\sum_{i=1}^c R^{-d_i}
	\end{equation}
	and this bound is strict if $\ell>0$.
\end{lemma}
\begin{proof}
	We use induction over the events (interruptions and completions).
	For an event at time $t$, let $\ell(t)$ be the number of interruptions so far, $c(t)$ the number of completions, and let $d_i(t)$ be the number of interruptions following the $i$th completion up until time $t$. Then we claim
	\[
	t\le U:=n_{\ell(t)}(a)+\sum_{i=1}^{c(t)} R^{-d_i(t)}.
	\]
	For $t=a$ we have $n_{0}(a)=a$ and $c(t)=0$, so the bound holds with equality.
	For the induction step, consider the two possible events at the next time.
	\begin{description}
		\item[completion] This takes place no later than time $U+1$ by induction. For the new completion we have $d_i(t)=0$, contributing 1 to the sum; all other counts remain unchanged. Hence the value of $U$ increases by exactly 1, proving the bound for this case.
		\item[interruption] Let this happen at time $t'<q(t)\le q(U)$. Since $q(x+y)= q(x)+y/R$, we get
		\[
		q(U) = n_{\ell(t)+1}(a) + \sum_{i=1}^{c(t)} R^{-(d_i(t)+1)}.
		\]
		This is the desired bound because $\ell(t')=\ell(t)+1$ and the number of interruptions following each completion has increased by 1. In this case we get a strict inequality which is maintained in later steps.\qedhere 
	\end{description}
\end{proof}

\begin{property}
\label{prop:lightest}
    The critical job is the lightest job in $S$.
\end{property}
\begin{proof}
    Suppose not. Let a job be \emph{heavy} if its weight is at least $1/R$ that of the critical job.
    We number the jobs in $S$ according to the order in which \rw completes them and starting from 0 for the last job in $S$ that is lighter than the critical job (any jobs in $S$ before job 0 have negative indices). 
    Then after job 0, all jobs in $S$ up to and including the critical job $z$ have weight at least 1. 
    Let $H$ be the set of these jobs in $S$, and let $k=|H|$. All jobs in $H$ are heavy, and
\[
	\rw(\ceone) = s_0+k+1,
\]
where $s_0$ denotes the time at which \rw starts job 0 (for the last time). Since \rw runs job 0 before any job in $H$, all jobs in $H$ arrive after time $s_0$ by Observation \ref{obs:simple}. and they are all at least as heavy as the critical job. We have 
\[
    \opt(\ceone)\ge s_0+k.
\] 
If $I$ is a counterexample, then $1/(s_0+k)>R-1$, so $s_0+k < 1/(R-1)$ and also  $\opt(\ceone) < 1/(R-1)$.

If no jobs from $H$ arrive before time $(s_0+2)/R-1$, then
\[
\opt(\ceone)\ge\frac{s_0+2}{R}+k-1
\] 
and $\rw(\ceone) = s_0+k+1 \le s_0+2+R(k-1) \le 
R\opt(\ceone),$ a contradiction. 
Therefore some job in $H$ arrives before $(s_0+2)/R-1$ (but after $s_0$). Since it does not interrupt job 0, we have $w_0\ge 1/\Delta$, so job 0 is heavy.

We have that \opt completes every heavy job no later than the last job of $H$. Otherwise, $\opt(\ceone) \ge (s_0+k+1)/R=\rw(\ceone)/R$, a contradiction. 
Since $\opt(\ceone) < 1/(R-1)<4$, \opt finishes the last job of $H$ before time 4. This means that there are at most three heavy jobs in the entire input including job $0$ and all jobs in $H$, implying $k\le 2$. 
In the calculations below, we use $\Delta^3\le R$ which implies $1/\Delta \ge \Delta/R$ and $1/\Delta^2 \ge \Delta/R$.

If $k=2$, the only heavy jobs are job 0 and the two jobs in $H$. Job 0 was not interrupted before time $s_0$, as this would have required a heavy job to arrive before time $s_0$. If job 0 arrives at time $s_0$, then $\opt(\ceone)\ge s_0+3$ since by Observation \ref{obs:simple} no job in $H$ arrives before job 0. This is a contradiction because $\rw(\ceone) = s_0 + 3$.
If job 0 starts after a non-heavy job $g$ completes, $s_0\ge1$ and $w_0\ge1/\Delta \ge \Delta/R > \Delta w_g$, so we must have $r_0\ge q(s_0-1)$ and $\opti\ge q(s_0-1)+3$, so $R\opti\ge s_0+1+2R\ge s_0+3=\rw(I)$.

If $k=1$, then $\rw(\ceone)=s_0+2$. In the optimal schedule, all heavy jobs (in particular job 0) precede or are equal to job 1. If job 0 arrives at time $s_0$, then $\opti\ge s_0+2=\rw(\ceone)$. Otherwise we consider the maximal interval $A$ in which heavy jobs are processed ending with the completion of job 0, and write $A=[a,s_0+1]$.

\emph{Claim 1:} If $A$ starts on a completion of a non-heavy job $j$, any job of weight at least $\Delta/R$ that is still unfinished when $A$ starts arrives at or after time $q(a-1)$.

\emph{Proof:} Job $j$ started at time $a-1$ and has weight less than $1/R$, so this follows from Observation \ref{obs:running} and $(\Delta/R)/(1/R)=\Delta$.

\emph{Claim 2:} If $A$ starts on an arrival and not on a completion, any job of weight at least $\Delta/R$ that is still unfinished when $A$ starts arrives at or after time $a$.

\emph{Proof:} If the machine was idle at time $a$, any such job $x$ certainly arrives at or after time $a$. If there was an interruption at time $a$, this was an interruption of a non-heavy job by definition of $A$; the fact that that job was still running until time $a$ and its cutoff was not yet reached (because it got interrupted at time $a$) implies that $x$ did not arrive before time $a$ by Observation \ref{obs:running}.

\bigskip 
Before job 1 there are at most two heavy jobs, namely job 0 and possibly a job $p$. $A$ therefore has one of the following three forms. Note that we have already excluded the case in which job 0 arrived at time $s_0$. 
\begin{enumerate}
	\item Job 0 alone. Then $a=s_0$ and a non-heavy job $g$ completes at time $s_0$. Then job 0 and job 1 arrived at or after time $q(s_0-1)$, as $w_0\ge1/\Delta \ge \Delta/R > \Delta w_g$ and $w_1=1$, so $R\opti\ge Rq(s_0-1)+2R\ge s_0+1+R\ge \rw(\ceone)$.
	\item Job $p$ which completes and is followed by job 0. Then $s_0=a+1$. If $r_0\ge q(a)$, then $\opti\ge q(a)+2=(a+2)/R-1+2$, so $R\opti\ge a+2+R\ge \rw(\ceone)$.
	Otherwise $w_p\ge w_0/\Delta \ge 1/\Delta^2 \ge \Delta/R$. If $p$ arrived at time $a$, we get $\opti\ge a+3 = \rw(\ceone)$ by applying Claim 2 to jobs 0, 1 and $p$. Else, we get $\opti\ge q(a-1)+3$ and $R\opti\ge \rw(\ceone)$ using Claim 1 on these jobs.
	\item Job 0 interrupted for $p$, which then completes and is followed by the final run of job 0. Then $s_0=s_p+1$ and $\rw(\ceone)=s_p+3 < q(a)+3$. If $A$ starts on an arrival and not on a completion, then $\opti\ge a+3$ by Claim 2, so
	\[
		\frac{\rw(\ceone)}{\opti}<\frac{q(a)+3}{a+3} \le \frac{2/R+2}{3}<R.
	\]
	Otherwise $a\ge1$ and $\opti\ge q(a-1)+3$ by Claim 1 and we get
	\[
		\frac{\rw(\ceone)}{\opti}<\frac{(a+2)/R+2}{(a+1)/R+2}=\frac{a+2+2R}{a+1+2R}=
		1+\frac{1}{a+1+2R}<R.\qedhere
	\]
\end{enumerate}
\end{proof}

Given Property 1, from this point on we simply again number the jobs in $S$ according to their order in $S$, with the first job in $S$ having index 1 and starting (for the last time) at time $s_1$. We define $k=|S|$, so we have 
\[\rw(I)=s_1+k.\]
\begin{figure}[h]
    \centering    
    \resizebox{\columnwidth}{!}{
 \tikzset{every picture/.style={transform shape}}
\centering
\begin{ganttchart}[
inline,
milestone/.append style={fill=black,  scale=.22 ,inner sep=0pt, shape=circle},
canvas/.style={draw=none},progress label text={}
]{2}{15}
\ganttbar[bar height=1,bar/.append style={ line width=1pt}]{\large{}}{4}{7}
\ganttbar[progress=100,
   bar height=1,
  bar/.append style={fill=palegray, line width=1pt},   
  bar incomplete/.append style={fill=white}]{\large{$1$}}{9}{12}
\ganttbar[progress=100,
   bar height=1,
  bar/.append style={fill=palegray, line width=1pt},   
  bar incomplete/.append style={fill=white}]{\large{$2$}}{13}{16}
\ganttbar[progress=100,
   bar height=1,
  bar/.append style={fill=palegray, line width=1pt},   
  bar incomplete/.append style={fill=white}]{\large{$k$}}{25}{28}
\ganttbar[bar height=1,bar/.append style={ line width=1pt}]{\large{}}{29}{32}
\ganttbar[bar height=1,bar/.append style={ line width=1pt}]{\large{}}{33}{36}
\ganttbar[
  progress=100,
   bar height=1,
  bar/.append style={fill=palegray, draw=none, line width=1pt},   
  bar incomplete/.append style={fill=white} 
]{}{17}{24}
 \begin{scope}[shift={(0,-1.4)}] 
 \draw[thick,->] (0,0) -- (19.3,0) node[anchor=north] {\large{time}};

    \draw[dashed,dash pattern=on 5pt off 4pt, line width=1pt] (0,1.1) -- (1,1.1);
    \draw[dashed,dash pattern=on 5pt off 4pt, line width=1pt] (0,0.1) -- (1,0.1);
    \node[text width = 0.5 cm] at (0.5,0.6) {...};    
    \node[text width = 0.5 cm] at (9.5,0.55) {...};
    \draw[dashed,dash pattern=on 5pt off 4pt, line width=1pt] (7.5,1.1) rectangle (11.5,1.1);
\draw[dashed,dash pattern=on 5pt off 4pt,line width=1pt] (7.5,0.1) rectangle (11.5,0.1);

    \draw[dashed, dash pattern=on 5pt off 4pt, line width=1pt] (17.5,1.1) -- (19,1.1);
    \draw[dashed,dash pattern=on 5pt off 4pt, line width=1pt] (17.5,0.1) -- (19,0.1);
    \node[text width = 0.5 cm] at (18.5,0.55) {...};
\end{scope}

\end{ganttchart}
    }
    \caption{This figure illustrates the set $S$ with jobs 1 up to $k$. Job $k$ is the critical job, and the jobs after $k$ are not in $S$ and are lighter.}
    \label{fig:setS}
\end{figure}

Note that there may be additional jobs that are run consecutively without interruption immediately after $S$, but not immediately before $S$. 
Let $r_S$ be the earliest arrival time of a job in $S$. 
\begin{property}
    \label{prop:early}
    We have $r_S<{(s_1+k)}/R-k$. 
\end{property}
\begin{proof}
    Suppose \opt starts running the first job in $S$ at time $(s_1+k)/R-k$ or later. Then 
    $\opt(I)\ge (s_1+k)/R$, so $I$ is not a counterexample.
    \qedhere
\end{proof}

\begin{property}[Properties of time $s_1$]
    \label{prop:1}
    No job completes at time $s_1$. Job 1 in $S$ arrives at time $s_1$. If a job in $S$ arrives before $s_1$, it is lighter than job 1. 	At time $s_1$, each unfinished job with weight at least $1/\Delta$ is in $S$.
\end{property}
\begin{proof}
    The first statement follows from the definition of $S$. It means that job 1 did not wait for another job to complete before starting. Since \rw also does not idle the machine if there is a job available, it starts job 1 immediately when it arrives, at time $s_1$.  The third statement follows from Observation \ref{obs:simple}.
    
    If the last statement does not hold for some job, its weighted completion time is at least $(\rw(\ceone)+1)/\Delta > \rw(\ceone)$, since $\rw(\ceone)<T+9<1/(\Delta-1)$. This would make this job critical, which is a contradiction.
    \qedhere
\end{proof}

\begin{property}
\label{prop:h}
    A job $h$ was interrupted and then job 1 starts at time 
    \[
        k(R-1)<s_1<T, \text{ and } 2\le k\le9.
    \] 
\end{property}
\begin{proof}
    If \opt starts its first job from $S$ at time $\frac{s_1}R-k(1-\frac1R)$ or later, then 
    $I$ is not a counterexample, as $|S|=k$. 
    Hence, \opt starts at least one job in $S$ at some time $s'<{(s_1+k)}/R-k<s_1$, implying that at least one job in $S$ arrives at or before time $s'$. 
    Since job 1 arrives at time $s_1$ by Property \ref{prop:1}, this means $k\ge2$.
    From $(s_1+k)/R-k>s'\ge 0$ we get $s_1>k(R-1)$.
    
    Since \rw does not complete any job in $S$ before time $s_1$ but also does not idle if at least one job is available, \rw does not idle in the interval $(s',s_1]$.
    By Property \ref{prop:1}, \rw does not complete a job at time $s_1$.
    Since \rw was not idle at time $s_1$, \rw interrupted some job $h\ne1$ at time $s_1$. 
    We get $s_1 < \frac{2-R}{R-1}$ by the rules of the algorithm. From $(R-1)k < s_1 < \frac{2-R}{R-1}$ we get $k\le 9$. \qedhere   
\end{proof}

\begin{property}
	\label{prop:onlyS}
	Let $0\le a<s_1$. Suppose $[a,s_1]$ is an interval in which \rw is not idle and in which no completions take place. If some job in $S$ is already waiting at time $a$ or arrives at time $a$, only jobs in $S$ are started in $[a,s_1]$.
\end{property}
\begin{proof}
	Under the assumptions, a job $j$ not in $S$ that is started in $[a,s_1]$ remains unfinished at time $s_1$. Since it is not in $S$ and the critical job is the lightest job in $S$, $j$ is lighter than all jobs in $S$. This contradicts that $j$ started when a job in $S$ was already waiting.
\end{proof}	
\begin{property}
	\label{obs:arrival}
	If there is no completion in $[r_S,s_1]$, then the first arriving job in $S$ starts on arrival at $r_S$, and $s_1<n_{k-1}(r_S)$.
\end{property}
\begin{proof}
	If not, there is a job $j$ running at time $r_S$ that has weight below 1 and is not in $S$. The time $r_S$ must be before its cutoff, else $j$ completes in $[r_S,s_1]$. Hence if the first arriving job in $S$ does not cause $j$ to be interrupted, we have $w_j\ge 1/\Delta$, contradicting Property \ref{prop:1}. 
    By Property \ref{prop:onlyS}, only jobs in $S$ can cause interruptions in $[r_S,s_1]$. Since $S$ contains $k$ jobs and a job can cause an interruption only when it arrives, there are at most $k-1$ interruptions in the interval $(r_S,s_1]$.
    The bound on $s_1$ follows from Lemma \ref{lem:s1bound}. 
\end{proof}

To assist in checking the following calculations, we provide the following table.
We also note that $q(0)=2/R-1\approx0.570$.

\begingroup
\setlength{\tabcolsep}{5pt}
\begin{tabular}{ccccccccc}
	$m$      & 1     & 2     & 3     & 4     & 5     & 6     & 7     & 8     \\
	\hline
	$n_m(0)$ & 0.570 & 1.017 & 1.368 & 1.644 & 1.860 & 2.030 & 2.163 & 2.268 \\
	$n_m(1)$ & 1.355 & 1.633 & 1.852 & 2.024 & 2.158 & 2.264 & 2.347 & 2.412 \\
	$n_m(2)$ & 2.140 & 2.249 & 2.336 & 2.403 & 2.456 & 2.498 & 2.531 & 2.556 \\
\end{tabular}
\endgroup

\begin{property}
	\label{prop:nocomp}
	\rw does not complete any job before time $s_1$.
\end{property}
\begin{proof}
	Let $p$ be the last job completed before $s_1$, let its final attempt start at $v$, and write $a=v+1$. If $r_S\ge a$, Property \ref{obs:arrival} gives $s_1<n_{k-1}(r_S)$ with $r_S\ge1$. 
    Using (\ref{eq:nmt}) we have 
	\begin{align*}
		\frac{n_{k-1}(r_S)+k}{r_S+k}
		& = \frac{R^{1-k} (r_S + k) + (1-R^{1-k})(T+k)}{r_S+k}
		  = R^{1-k} + \frac{(1-R^{1-k})(T+k)}{r_S+k}.
	\end{align*}
	This is decreasing in $r_S$. Setting $r_S=1$ we get
	\begin{equation}
		\label{eq:onecase}
	\frac{\rw(\ceone)}{\opti} \le \frac{n_{k-1}(1)+k}{1+k}<1.171.
	\qquad\text{Worst case: }k=5
	\end{equation}
	
	Assume henceforth $r_S<a$. This means that at time $a$ a job of $S$ is waiting. Between $a$ and $s_1$ there are $m$ interruptions and no completions, where $1\le m\le k-1$. By Property \ref{prop:onlyS} and Lemma \ref{lem:s1bound}, $s_1<n_m(v+1)$. 
	
	\proofsubparagraph*{Case 1: $w_p<1/\Delta$.}
	By Observation \ref{obs:running}, $r_S\ge q(v)$. 
	Using $v+1 = Rq(v)+R-1$  and (\ref{eq:nmt}),
	\begin{align*}
		n_m(v+1)+k &= R^{-m}(Rq(v)+R-1)+(1-R^{-m})T+ k\\
		&= R^{1-m} q(v)+(1-R^{1-m})T+R^{-m}+k\\
		&= R^{1-m} (q(v)+k)+(1-R^{1-m})(T+k)+R^{-m}.
	\end{align*}
	The second equality uses $(R-1)T=2-R$.
	If $m\le k-2$, then 
	\[
		\frac{\rw(\ceone)}{\opti} \le \frac{n_{k-2}(v+1)+k}{q(v)+k}
		= R^{3-k} + \frac{(1-R^{3-k})(T+k)+R^{2-k}}{q(v)+k}.
	\]	
	This is decreasing in $v$, and we conclude 
	\[
		\frac{\rw(\ceone)}{\opti} \le \frac{n_{k-2}(1)+k}{q(0)+k}<1.233.
		\qquad\text{Worst case: }k=4
	\]	
	If $m=k-1$, the chain contains all jobs of $S$. Its initial job is the critical job $z$, and all other jobs have weight greater than $\Delta$ and arrive after $a=v+1$. If $z$ is last among $S$ in OPT, then exactly as in (\ref{eq:onecase})
	\[
		\frac{\rw(\ceone)}{\opti} \le \frac{n_{k-1}(v+1)+k}{v+1+k} < 1.171.
		\qquad\text{Worst case: }k=5
	\]
	Otherwise as before 
	\[
		\frac{\rw(\ceone)}{\opti} \le \frac{n_{k-1}(v+1)+k}{\Delta(r_S+k)} 
		\le \frac{n_{k-1}(v+1)+k}{\Delta(q(v)+k)}
		\le \frac{n_{k-1}(1)+k}{\Delta(q(0)+k)}
		< 1.214.
	\]
	
	\proofsubparagraph*{Case 2: $1/\Delta\le w_p<1$.}
	All jobs of $S$ arrive at or after $v$. If $p$ starts on arrival, all jobs in $\{p\}\cup S$ have release times at least $v$ and weights at least $1/\Delta$. Thus $\opti\ge(v+k+1)/\Delta$ and as in (\ref{eq:onecase}) we get
	\[
		\frac{\rw(\ceone)}{\opti} \le \frac{n_{k-1}(v+1)+k}{(v+1+k)/\Delta}
		\le 
		\frac{n_{k-1}(1)+k}{(1+k)/\Delta}
		<1.26.
		\qquad\text{Worst case: }k=5
	\]	
	Otherwise a job $g$ completes at $v\ge1$ and $p$ starts immediately after it. By Observation \ref{obs:running}, $r_p\ge q(v-1)$ or $w_g\ge w_p/\Delta\ge1/\Delta^2$. 
	
	If $r_p\ge q(v-1)$, all jobs in $\{p\}\cup S$ arrive at or after $q(v-1)$, giving $\opti\ge(q(v-1)+k+1)/\Delta$ and a ratio below $1.248$. (Worst case: $k=2,v=1$.)
	
	If $w_g\ge w_p/\Delta\ge1/\Delta^2$, then if both $g$ and $p$ finish before the last job of $S$ in OPT, $\opti\ge k+2$. If either finishes after that last job, then $\opti\ge(v+k+1)/\Delta^2$, since all $k$ jobs of $S$ arrive at or after $v$. Hence $\opti\ge\min\{k+2,(v+k+1)/\Delta^2\}$. Using $s_1<T$ for the first denominator and $s_1<n_{k-1}(v+1)$ for the second gives 
	\[
		\frac{\rw(\ceone)}{\opti} \le \frac{s_1+k}{\min\{k+2,(v+k+1)/\Delta^2\}}<1.224.
		\qquad\text{Worst case: }k=5
	\]
	
	\proofsubparagraph*{Case 3: $w_p\ge1$.}
	Put $H=S\cup\{p\}$ and $r_H=\min_{j\in H}r_j$. Suppose first that $p$ is the only job completed in $[r_H,s_1]$. The first job of $H$ starts on arrival at $r_H$. To see this, a lighter running job (which is by definition of $r_H$ not in $H$) could neither complete before $s_1$ (because it is not $p$) nor reject the arrival and remain unfinished with weight at least $1/\Delta$ by Property \ref{prop:1}. 
	
	From $r_H$ to $s_1$ therefore the machine processes only jobs of $H$ because $r_S<v+1$. There are at most $k$ interruptions, because every interruption uses a new arrival. The completion of $p$ is followed by at least one interruption (Property \ref{prop:h}).
	If there are at most $k-1$ interruptions, \eqref{eq:events} yields $s_1<n_{k-1}(r_H)+1/R$. Together with $\opti\ge r_H+k+1$, this gives $\rw(\ceone)/\opti<1.239$, where the highest bound is achieved for $k=5$.
	
	If there are exactly $k$ interruptions, every job of $H$ except the first arrives as an interruptor while a job of weight at least $1$ runs, so every such job has weight greater than $\Delta$. The first arriving and running job is therefore $z$, with $p$ to come later. If $z$ is not last among $H$ in OPT, then $\opti\ge\Delta(r_H+k+1)$, giving 
	\[
		\frac{\rw(\ceone)}{\opti} < \frac{n_k(r_H)+1/R+k}{\Delta(r_H+k+1)}
		\le \frac{n_k(0)+1/R+k}{\Delta(k+1)}<1.198.
		\quad\text{Worst case: }k=3
	\]
	
	If $z$ is last, let $u$ be the first interruption time. All other jobs of $H$ arrive at or after $u$, giving $\opti\ge u+k+1$. From $u$ there are only $k-1$ interruptions and the one completion, so the preceding bound below $1.239$ applies.
	
	The remaining possibility is another completion of a job $g$ after $r_H$ (and before job $p$). There can be at most one such job as $s_1<T<3$.
	If $w_g\ge1/\Delta$, the $k+2$ jobs give $\opti\ge(k+2)/\Delta$, and 
	\[
		\frac{\rw(\ceone)}{\opti}<\frac{\Delta(T+k)}{k+2}<1.251.
		\qquad\text{Worst case: }k=2
	\] 

	If $w_g<1/\Delta$, let its final attempt start at $b$ and finish at $c=b+1$. Since $r_H<c$, a job of $H$ is waiting at $c$. All jobs of $H$ arrive at or after $q(b)$, as any earlier one would interrupt $g$'s final attempt. From $c$ to $s$ there are at most $k$ interruptions, and the one completion of $p$, followed by an interruption. Thus $s_1<n_k(b+1)+1/R$ and $\opti\ge q(b)+k+1$, giving  
	\[
		\frac{\rw(\ceone)}{\opti}\le\frac{n_k(b+1)+1/R+k}{q(b)+k+1}<\lbratio.
		\qquad\text{Worst case: }k=2
	\]
	All cases contradict $\rw(\ceone)>R\opti$.\qedhere 
\end{proof}

\begin{figure}
\begin{tikzpicture}

\draw (0, 0) -- (0, 1);
\draw (0, 0.5) -- (1, 0.5);

\draw (1, 0) rectangle (11, 1);

\draw (3, 0) -- (3, 1);
\draw (4.5, 0) -- (4.5, 1);
\draw (5.5, 0) -- (5.5, 1);

\node at (2, 0.5) {$4$};
\node at (3.75, 0.5) {$3$};
\node at (5, 0.5) {$2$};
\node at (7.75, 0.5) {\small Job 1, then 2,3,4};

\node[below] at (0, 0) {$0$};
\node[below] at (1, 0) {$r_S$};
\node[below] at (3, 0) {$t_1$};
\node[below] at (4.5, 0) {$t_2$};
\node[below] at (5.5, 0) {$t_3=s_1$};

\end{tikzpicture}
    \caption{The final shape of the hypothetical counterexample, here shown for $k=4$ and $m=k-1$. The first job in $S$ that arrives is job 4, and it arrives at time $r_S$. Then there are interruptions for increasingly heavier jobs, until time $s_1$ when job 1 starts and is not interrupted. Jobs are then completed in order of decreasing weight. We have $t_i< n_i(r_S)$ for $i=1,2,3$.}
    \label{fig:placeholder}
\end{figure}

\begin{theorem}
	\label{thm:rw}
	The RW algorithm is exactly $R$-competitive.
\end{theorem}
\begin{proof}
By Property \ref{prop:nocomp}, there are no completions before $s_1$. By Property \ref{obs:arrival}, the first job of $S$ starts on arrival at $r_S$. From $r_S$ to $s_1$ there is a chain of $m$ interruptions, where $1\le m\le k-1$, all involving jobs of $S$.

If $m\le k-2$, then $s_1<n_{k-2}(r_S)$ and $\opti\ge r_S+k$. We obtain
\begin{equation}\label{eq:final}
	\frac{\rw(\ceone)}\opti<\frac{n_{k-2}(r_S)+k}{r_S+k}\le\frac{n_{k-2}(0)+k}{k}\le R.
\end{equation}
The last inequality holds for $2\le k\le9$, with exact equality for $k=6$. 
That is, we claim that $(n_4(0)+6)/6=R$, or equivalently $n_4(0)=6(R-1)$.
As $n_m(t)=R^{-m}t+(1-R^{-m})T$, we have $n_4(0)=(1-R^{-4})\frac{2-R}{R-1}$.
We conclude that indeed $n_4(0)-6(R-1)=0$ because 
\begin{align*}
R^4\bigl(n_4(0)-6(R-1)\bigr)
&=\frac{(R^4-1)(2-R)}{R-1}-6R^4(R-1)\\
&=(R^3+R^2+R+1)(2-R)-6R^4(R-1)\\
&=-6R^5+5R^4+R^3+R^2+R+2=0.
\end{align*}
This shows that the analysis of \rw is tight.

If $m=k-1$, every job in $S$ belongs to the chain. Its initial job is $z$, and all other jobs have weight greater than $\Delta$. If $z$ is not last among $S$ in OPT, then $\opti\ge\Delta(r_S+k)$, and 
\[
	\frac{\rw(\ceone)}{\opti}\le \frac{n_{k-1}(r_S)+k}{\Delta(r_S+k)}<1.248.
	\qquad\text{Worst case: }k=4,r_S=0
\]

If $z$ is last among $S$ in OPT, let $u$ be the first interruption time. The other $k-1$ jobs all arrive at or after $u$, and must precede $z$ in OPT. Hence $\opti\ge u+k$. There are only $k-2$ interruptions after $u$, so $s_1\le n_{k-2}(u)$. The same calculation as \eqref{eq:final} gives $\rw(\ceone)/\opti\le R$. For $k=2$, the ratio is 1 as $n_0(u)=u$.\qedhere
\end{proof}

In the next theorem, we complement \Cref{thm:rw} with a lower bound, leaving a small gap for further improvement.
\begin{theorem}
\label{thm:lb12344}
Any deterministic online algorithm for the problem $1|r_j, \text{online}, p_j=1 ,\text{restart} | W C_{\max}$ is at least $R_2\approx \lbratio$-competitive which is the positive real root of the equation $8x^4-2x^3-15=0$.
\end{theorem}

\begin{proof}
    For a contradiction, suppose that there exists an online algorithm $\alg$ with a competitive ratio of less than $R_2$. We use the notation $c_i$ for the completion time of job $i$ by $\alg$.
    We consider the following job instance. At time $r_1=0$, job 1 with $w_1=1$ arrives. 
    Then job 2 arrives with $w_2=\frac{4}{3}$ at time $r_2 = 3/{R_2}-2\approx0.423$.
    At this point the optimal schedule is 2 followed by 1 for a cost of $3/R_2$.
    
    \proofsubparagraph*{Case 1:} \alg completes job 1 before starting job 2. If $c_1\ge5/4$, no further jobs arrive and $\frac43(c_1+1) \ge 3 = R_2\opt(I)$.
    Else, a final job 3 with weight $1+R_2/3 \approx 1.413$ arrives at time $1 + r_2 = 3/{R_2} -1\approx1.423$. The following schedule is feasible: job 2 is completed at time $1 + r_2$, job 3 is completed at time $2 + r_2$ and job 1 is completed at time $3 + r_2$.
    We have $\opt(I) \le 3 + r_2 = 3/{R_2} +1\approx3.423$ as $(2+r_2)(1+R_2/3)=
    (1+R_2/3)\cdot3/R_2 = 3/R_2+1 = 3+r_2$. 
    
    If \alg completes job 2 before job 3, then \[\alg(I)\ge 3\left(1+\frac{R_2}3\right)=3+R_2 = R_2\left(\frac{3}{R_2}+1\right).\] 
    If \alg completes job 3 before job 2, then \[\alg(I)\ge \frac43\left(\frac{3}{R_2}+1\right)>R_2\left(\frac{3}{R_2}+1\right).\] 

    \proofsubparagraph*{Case 2:} \alg starts job 2 before finishing job 1, at time $s_2$. If $s_2\ge1$, the input stops and $\alg(I)\ge3=R_2\opt(I)$.
    Else, a (different) job 3 with $w_3=\frac{5}{2R_2}$ arrives at time $r_3$. Here $r_3=4R_2-4\approx0.952$ if $s_2<4R_2-4$, else $r_3=(s_2+1)/2$.
    
    The optimal schedule has value at most $5/R_2>4$ by using the order $3,2,1$, as the weighted completion times are \[w_3(r_3+1)<2w_3=\frac5{R_2},\quad \frac43(r_3+2)<4,\text{ and } r_3+3<4.\]
    When $s_2<4R_2-4$, we have $r_3=4R_2-4$ and $\opt(I)=4R_2-1\approx 3.952$ as $w_3(4R_2-3)<3.943$ and $\frac43(4R_2-2)<3.937$.
    
    If \alg completes no job before time 3, $\alg(I)\ge5$ and we are done.
    
    \proofsubparagraph*{Case 2a:}
    $\alg$ completes job 1 or 2 before job 3. No more jobs arrive. 
    \alg completes its first job not before time $s_2+1$, so $c_3\ge s_2+2$ and $\alg(I)\ge \frac{5}{2R_2}(s_2+2)$. 
    
    If $s_2<4R_2-4$, then $\opt(I)\le 4R_2-1$ and $\alg(I)\ge 3\frac{5}{2R_2^2}=R_2(4R_2-1)$. This equality follows from the fact that $R_2$ is a solution of $8x^4-2x^3-15=0$.
    
    If $s_2\ge 4R_2-4$, we use that $\opt(I)\le 5/R_2$ and 
    we have $\alg(I)/\opt(I)\ge (4R_2-2)/2>R_2$.
    
    \proofsubparagraph*{Case 2b:} $\alg$ completes job 3 first, before time 3. A final fourth job 4 with $w_4=\frac{5}{4R_2}\approx1.0096$ arrives at time $r_4=3$. A feasible schedule $s$ would be the job order $1,3,2,4$ with $WC_{s} = 5/R_2$. If $\alg$ completes job 4 last, its completion time is at least $r_3+4\ge 4R_2$, so $\alg(I)\ge 5$. Else, some job completes after job 4 which is released at time 3, so again $\alg(I)\ge5$.\qedhere
\end{proof}

\section{Conclusion}
We studied the problem of online scheduling with restarts for minimizing the weighted makespan $WC_{\max}$. We presented two new lower bounds on the competitive ratio for deterministic algorithms with restarts, one in the general setting and another in the case where all jobs have identical processing times. We designed a deterministic online algorithm for unit processing times that improves the competitive ratio to better than \ubratio. This leaves a small gap to our lower bound for further improvement. Closing this gap appears to be nontrivial and is an interesting topic for future research.
This result may also lead the way to an improved result for general processing times.

\paragraph*{Acknowledgements}
The conference version of this paper~\cite{AJPSW26} was prepared without the help of AI. After the conference, several ideas for improvements were discussed with ChatGPT Astra. Eventually, it constructed a proof for one of these improvements based on the already existing proof structure. This proof was edited extensively by the authors. ChatGPT also found a small improvement to the lower bound construction.

\paragraph*{Disclosure of Interests}
The authors have no competing interests to declare that are relevant to the content of this article.

\bibliographystyle{plainurl}
\bibliography{bib}

\end{document}